\documentclass[aip,reprint]{revtex4-1}
\usepackage{graphicx}
\usepackage{siunitx}

\begin{document}


\title{A facile process for soak-and-peel delamination of CVD graphene from substrates using water} 



\author{Priti Gupta}
\altaffiliation{These authors contributed equally to this work.}
\affiliation{Department of Condensed Matter Physics and Materials Science, Tata Institute of Fundamental Research, Mumbai 400005, India
}

\author{Pratiksha D. Dongare}
\altaffiliation{These authors contributed equally to this work.}
\affiliation{Department of Condensed Matter Physics and Materials Science, Tata Institute of Fundamental Research, Mumbai 400005, India
}
\affiliation{Birla Institute of Technology and Science, Vidyavihar Campus, Pilani, Rajasthan 333031, India}

\author{Sameer Grover}
\affiliation{Department of Condensed Matter Physics and Materials Science, Tata Institute of Fundamental Research, Mumbai 400005, India
}

\author{Sudipta Dubey}
\affiliation{Department of Condensed Matter Physics and Materials Science, Tata Institute of Fundamental Research, Mumbai 400005, India
}

\author{Hitesh Mamgain}
\affiliation{WITec GmbH, Lise-Meitner-Str. 6, D-89081 Ulm, Germany}

\author{Arnab Bhattacharya}%
\email{arnab@tifr.res.in}%
\affiliation{Department of Condensed Matter Physics and Materials Science, Tata Institute of Fundamental Research, Mumbai 400005, India
}

\author{Mandar M. Deshmukh}%
\email{deshmukh@tifr.res.in}%
\affiliation{Department of Condensed Matter Physics and Materials Science, Tata Institute of Fundamental Research, Mumbai 400005, India
}



\date{\today}

\begin{abstract}
 We demonstrate a simple technique to transfer CVD-grown graphene from copper and platinum substrates using a soak-and-peel delamination technique utilizing only hot deionized water. The lack of chemical etchants results in cleaner CVD graphene films minimizing unintentional doping, as confirmed by Raman and electrical measurements. The process allows the reuse of substrates and hence can enable the use of oriented substrates for growth of higher quality graphene, and is an inherently inexpensive and scalable process for large-area production.
\end{abstract}


\maketitle 

Graphene, a monolayer honeycomb lattice structure of sp$^{2}$-bonded carbon atoms, has become a subject of great interest due to its extraordinary optical, mechanical, and  electronic properties \cite{geim_rise_2007,neto_electronic_2007,novoselov_roadmap_2012}.
Successful isolation of graphene by the mechanical exfoliation of highly oriented pyrolytic graphite (HOPG) has opened doors for new innovations in the field of nanoelectronics \cite{novoselov_electric_2004,zhang_fabrication_2005}. Since then many new methods have emerged to synthesize and isolate single to few-layer graphene \cite{novoselov_roadmap_2012} especially on large area substrates. These methods include reduction of graphite oxide \cite{stankovich_synthesis_2007}, ultrasonication of graphite \cite{hernandez_high-yield_2008}, synthesis on SiC substrate \cite{berger_ultrathin_2004}, and chemical vapor deposition (CVD) on metal substrates such as Ni \cite{yu_graphene_2008}, Cu \cite{li_large-area_2009}, Ru \cite{sutter_epitaxial_2008} and Pt \cite{kang_monolayer_2009,sutter_graphene_2009}.

Of all these methods, the low-pressure growth of graphene on Cu foils, in particular, is known to be advantageous in terms of controlled graphene size, number of layers and quality \cite{li_large-area_2009}. It has also been shown that better quality graphene can be grown on Cu(111) oriented grains \cite{wood_effects_2011}. Thus CVD graphene growth on Cu produces large areas of mostly monolayer graphene and is a promising way of producing large area graphene for practical nanoelectronics applications \cite{geim_graphene:_2009,novoselov_roadmap_2012}. To fully realize the advantages of the CVD graphene growth there must be a reliable method for transferring the graphene from metallic Cu substrates to more useful substrates like insulating substrates \cite{reina_large_2009}, flexible/stretchable substrates \cite{kim_large-scale_2009}, and transparent electrodes \cite{bonaccorso_graphene_2010,bae_roll--roll_2010}. Pt(111), as a substrate for CVD graphene growth, is also interesting because it has minimum effect on the physical properties of graphene due to its very weak graphene-substrate interaction. Further, it has been shown that the electronic structure of the graphene grown on Pt is nearly the same as that of the free standing graphene \cite{gao_epitaxial_2011}. In addition, Pt does not get oxidized easily like other metal substrates such as Cu.

Currently the processes used to transfer large-area and high-quality graphene synthesized on metal substrates require wet etching of the metal substrates \cite{li_large-area_2009,li_transfer_2009}. These processes trap ionic species between graphene and substrate interface which act as scattering centers and lead to degradation of the electrical properties of the devices fabricated on the graphene. Further, the etching process also results in loss of metal ultimately increasing the cost of the transfer process; this is especially true for precious metals and oriented single crystal substrates that are expensive. Electrochemical methods to transfer graphene without metal loss have been demonstrated but they involve chemicals like NaOH and are complex \cite{Wang2011electrochemical,gao_repeated_2012}. An intercalation method to transfer graphene from Pt to other substrates has also been shown, but for small
size graphene flakes only \cite{Ma}. To overcome these problems, we demonstrate, a novel facile method to transfer graphene from metal substrates (Cu and Pt) with hot deionized (DI) water without using any chemical etchants. This results in transferred graphene layers that are clean and show improved properties compared to graphene layers transferred using the conventional etching route.

DI water has potential use in transfer processes due to its capability to penetrate nanoscale hydrophobic-hydrophilic interfaces and separate them. Such methods have been used to selectively position\cite{bonaccorso_graphene_2010} and transfer\cite{schneider_wedging_2010} graphene flakes and other nanostructures using difference in affinity to water. In our method, we first coat the graphene layer with poly(methyl methacrylate) (PMMA) as a support material. PMMA has been used to support and transfer mechanically exfoliated graphene flakes\cite{bonaccorso_production_2012,reina_transferring_2008} and CVD-grown graphene to target substrates\cite{reina_large_2009}. We then exploit the differential interaction of water with the hydrophobic graphene \cite{bianco2012understanding, wang2009wettability} and the hydrophilic metal like Cu\cite{valette1982hydrophilicity} or Pt{\cite{gardner1977hydrophilic} to delaminate the graphene from the substrate used for CVD growth.

The most important difference in our DI water Soak-and-Peel Delamination (SPeeD) method and presently established methods to transfer graphene from Cu \cite{bonaccorso_production_2012,reina_transferring_2008,li_large-area_2009,levendorf_transfer-free_2009,kim_large-scale_2009,bae_roll--roll_2010,li_transfer_2009,lee_wafer-scale_2010,regan_direct_2010} and Pt substrates \cite{gao_repeated_2012} is that, our SPeeD method is very simple since it does not involve use of any chemical etchants and hence provides cleaner graphene. Additionally the metal substrate (Cu and Pt) is retained and can be reused --- an aspect desirable for industrial production. As our technique utilizes the difference in the interaction of the graphene  and substrates with water, this method can be extended to a larger class of  CVD substrates for a variety of applications. The SPeeD technique uses only DI water, hence  contamination due to ionic species can be significantly reduced ensuring that the electrical properties are not degraded as typically seen for graphene transferred via processes using chemical etchants to remove the Cu substrate.

\begin{figure}[!h]
\centerline{\includegraphics[width=100mm]{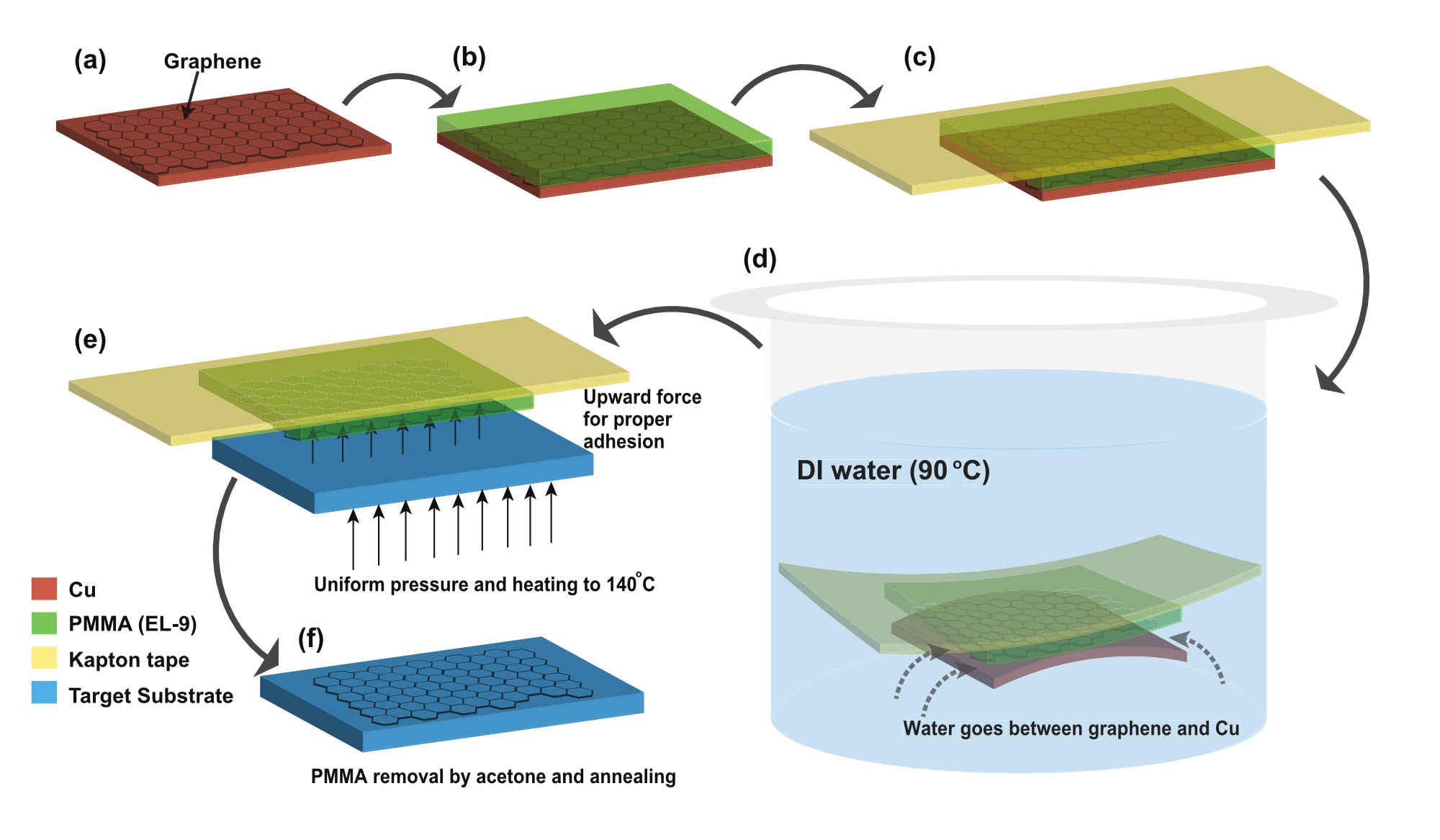}}
\caption{Schematic showing the steps of the DI water SPeeD graphene transfer method. (a) CVD graphene is grown on Cu substrate. (b) PMMA is spin-coated on the graphene grown on Cu/ Pt substrate. (c) Kapton tape is pressed on PMMA with a teflon roller. (d) The stack is immersed in DI water at \SI{90}{\degreeCelsius}. Water penetrates between graphene and Cu substrate and separates them. (e) Kapton tape with PMMA and graphene on it is pressed against the target substrate and heated for $40$~minutes at \SI{140}{\degreeCelsius}. (f) The PMMA on the target substrate is removed with acetone and RTA.
}
\label{fig:fig1}
\end{figure}

 CVD graphene was grown both in continuous\cite{li_large-area_2009} and island growth\cite{han_suppression_2012} modes on Cu and Pt substrates (details about the growth provided in Supporting Information section~I \cite{supplementary}). The SPeeD transfer process is schematically depicted in Figure~\ref{fig:fig1}. After the CVD growth of graphene on Cu foil (Figure~\ref{fig:fig1}(a)), we spin-coated the graphene with PMMA ($310$~nm thick resist (Microchem EL~$9$) at a speed of $3200$~rpm for $45$~s) followed by baking for $7$~minutes at \SI{175}{\degreeCelsius} (Figure~\ref{fig:fig1}(b)). To avoid crumpling of the resist after delamination and for easier handling, Kapton tape ($3$~M $5413$) is stuck on the resist and uniform pressure is applied using a teflon roller (Figure~\ref{fig:fig1}(c)).

The Cu foil, with the tape attached, is then immersed in a beaker of DI water maintained at \SI{90}{\degreeCelsius} for $2$~hours (Figure~\ref{fig:fig1}(d)). During this period DI water penetrates the graphene-Cu interface. Subsequently, the Kapton tape (to which the PMMA/graphene stack is attached) is slowly peeled away with tweezers leaving behind the Cu foil (see supporting movie~S1 which shows the key steps of our SPeeD process \cite{supplementary}).

\begin{figure*}[ht]
\centerline{\includegraphics[width=150mm]{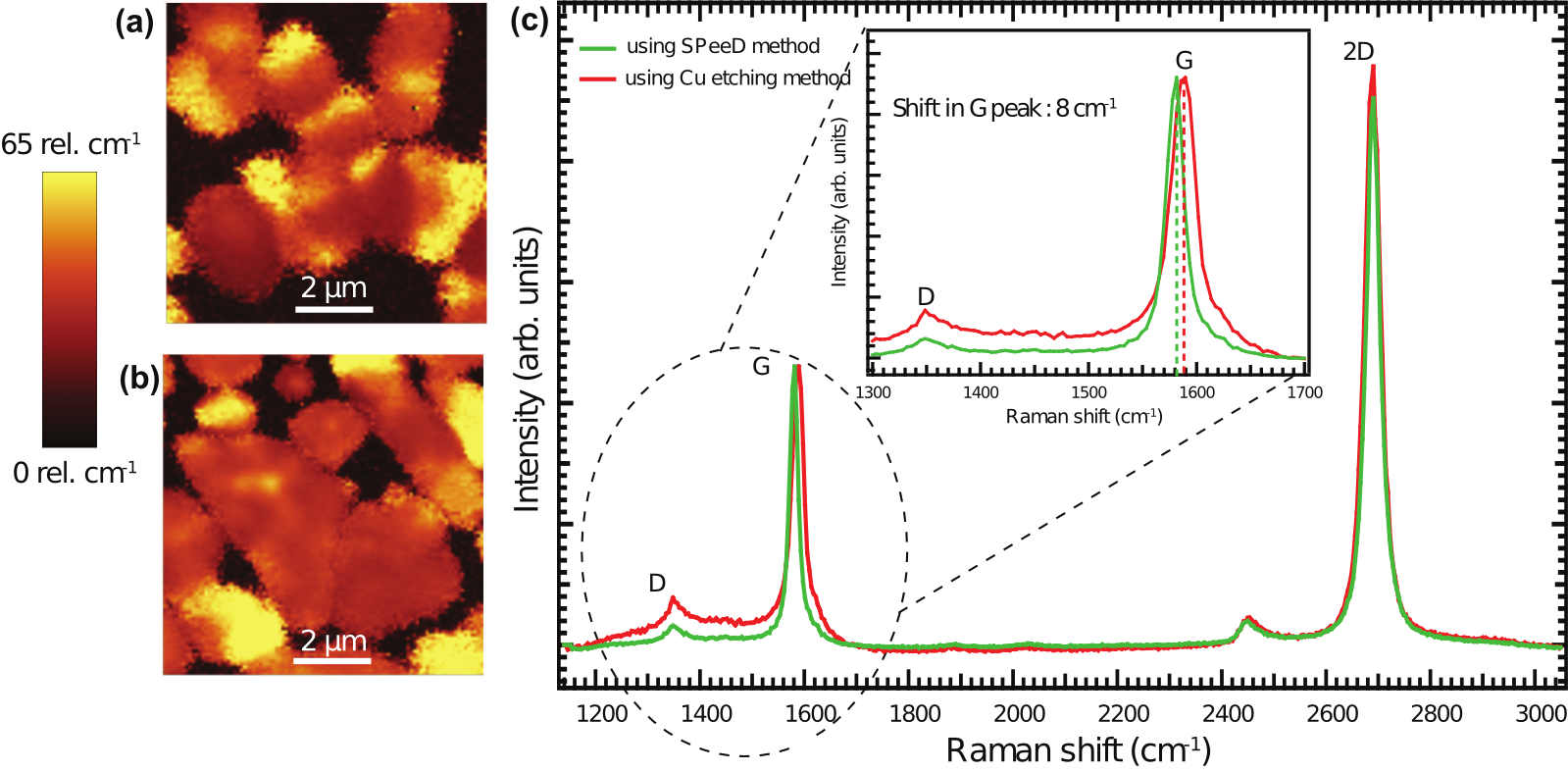}}
\caption{Comparative Raman spectroscopy of graphene transferred by etching Cu to release graphene and by using the SPeeD process. Integrated Raman mapping over an area of \SI{9}{\micro\metre}~$\times$~\SI{9}{\micro\metre} for the FWHM of the 2D peak of graphene transferred by (a) our SPeeD method using DI water (b) conventional etching of Cu by ammonium persulphate solution. (c) Comparison of the spatially averaged Raman scattering spectra of graphene transferred by the two different methods. Inset of (c) shows the blue shift of $8$~cm$^{-1}$ in the G peak for graphene transferred by Cu etching which indicates that it is p-type doped.}
\label{fig:fig2}
\end{figure*}

The target substrate, a $300$~nm thick SiO$_{2}$-coated p-type silicon wafer, was cleaned using oxygen plasma reactive ion etching to ensure better adhesion between graphene and the substrate. The Kapton tape, clamped to a glass slide, and the target substrate are brought into contact and heated for $40$~minutes at \SI{140}{\degreeCelsius} (Figure~\ref{fig:fig1}(e)). After allowing the sample to cool for $20$~minutes, the Kapton tape is detached from the glass slide and the stack is put in acetone to remove the EL~$9$ layer. Subsequently, rapid thermal annealing (RTA) (\SI{300}{\degreeCelsius} for 10 minutes and \SI{350}{\degreeCelsius} for 5 minutes in 100~sccm Ar) is done to remove any residual PMMA (Figure~\ref{fig:fig1}(f)), thus completing the transfer process.

We used the DI water based SPeeD method to transfer the graphene grown on Pt foils as well. With our technique, this can be easily done without curling of graphene and without using any chemical like NaOH that can lead to unintentional doping of graphene \cite{gao_repeated_2012}. The only difference from the process for releasing used for graphene on Cu is that a thicker resist layer was used for Pt compared to Cu and the resist was not baked. The remaining procedure for transfer of graphene from Pt is same as that for graphene on Cu.


To benchmark the quality of graphene transferred using our SPeeD based transfer process we transferred two graphene samples, grown on Cu foils from the same batch and in the same graphene growth run, by two different transfer methods. Though the quality of our CVD graphene is not as good as the best reported in the literature \cite{li_large-area_2009,li_large-area_2011,reina_large_2009}, the comparison of graphene samples grown in the same run under identical conditions but transferred by different processes should show the influence of the transfer process on the defect level, doping level and quality of graphene. The first sample was transferred onto an SiO$_2$-coated Si substrate using conventional etching of Cu with ammonium persulphate solution (details in Supporting Information section~II). The second sample was transferred to an identical substrate using our SPeeD method with DI water without using any etchant. We compare the two samples transferred using Raman spectroscopy measurements \cite{ferrari_raman_2013} and electrical transport measurements. The results of the comparison are discussed in the following sections.


Confocal Raman spectroscopy measurements were performed on both the samples using a WITec Alpha $300$R confocal Raman microscope. Figures~\ref{fig:fig2}(a) and (b) compare the Raman maps of the full width at half maximum (FWHM) of the 2D peak for the two graphene samples transferred by the two different methods. The spatially-averaged Raman spectra of graphene over a \SI{9}{\micro\metre}~$\times$~\SI{9}{\micro\metre} area for both samples are shown in Figure~\ref{fig:fig2}(c). To evaluate shift in peak positions, the spectra are aligned with reference to the Si substrate peak ($520$~cm$^{-1}$), and to compare relative intensity of the various features, the spectra are intensity normalized to the graphene G~peak value. The ratio of the intensity of the 2D peak to G peak in both the samples is comparable and has the value of $\sim$$2$, which indicates that the graphene is monolayer \cite{ferrari_raman_2013}.

The interesting observation is that the integrated Raman spectrum of the graphene transferred by the SPeeD method has a lower D peak intensity than that of the sample transferred by conventional Cu etching (Figure~\ref{fig:fig2}(c)). This suggests that the graphene transferred by the SPeeD process has less defects compared to the one transferred by conventional Cu etching.\cite{ferrari_raman_2007} Further, the inset (Figure~\ref{fig:fig2}(c)) shows that the FWHM of the G peak of the graphene transferred by the SPeeD process is narrower than that of the graphene transferred by Cu etching. This points to a  lower disorder in the SPeeD transferred sample.\cite{canccado2011quantifying} Additionally, the G peak is blue shifted \cite{das_monitoring_2008} in the graphene transferred by Cu etching. This indicates that the graphene transferred by Cu etching is p-type doped compared to the other sample \cite{das_monitoring_2008,ferrari_raman_2013}. This p-type doping can be attributed to charge impurities present in the Cu etchant or the presence of defects in the graphene. Thus, transferring graphene using our SPeeD method without any use of Cu etchant reduces the probability of graphene getting doped by ionic impurities. Transport measurements on two graphene samples (discussed later) grown using the same recipe but different transfer methods corroborate this observation about reduced doping in the SPeeD transferred samples.

The SPeeD method has been successfully applied to transfer CVD graphene grown on Pt foils as well. Raman measurements on the CVD graphene transferred via the SPeeD method from Pt foils to $300$~nm SiO$_{2}$-coated p-doped Si are shown in Figure~\ref{fig:fig3}.

\begin{figure}[ht]
\centerline{\includegraphics[width=90mm]{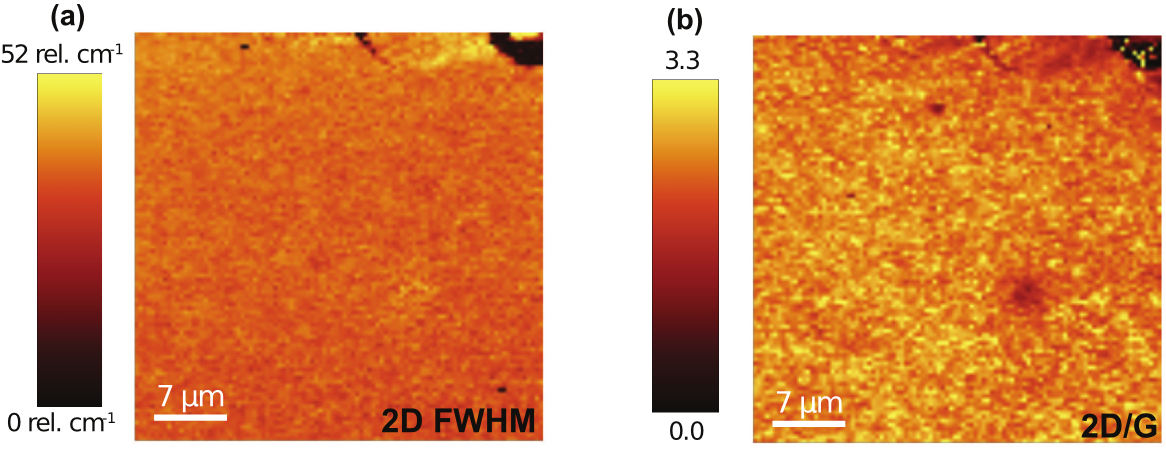}}
\caption{Raman measurement data for the CVD graphene grown on Pt and transferred using SPeeD method. Colorscale maps of (a) the width of the 2D peak (b) 2D/G peak intensity ratio across an area \SI{40}{\micro\metre}~$\times$~\SI{40}{\micro\metre} of graphene grown on Pt.   }

\label{fig:fig3}
\end{figure}


Figure~\ref{fig:fig4}(a) shows the optical image of the graphene transferred from Cu substrate by SPeeD method on $300$~nm SiO$_{2}$-coated p-type doped silicon target substrate. Electrical transport measurement data on two graphene samples grown under same conditions, but one transferred via conventional Cu etching and the other via our DI water SPeeD method are shown in Figure~\ref{fig:fig4}(b) and (c) respectively.
\begin{figure}[ht]
\centerline{\includegraphics[width=90mm]{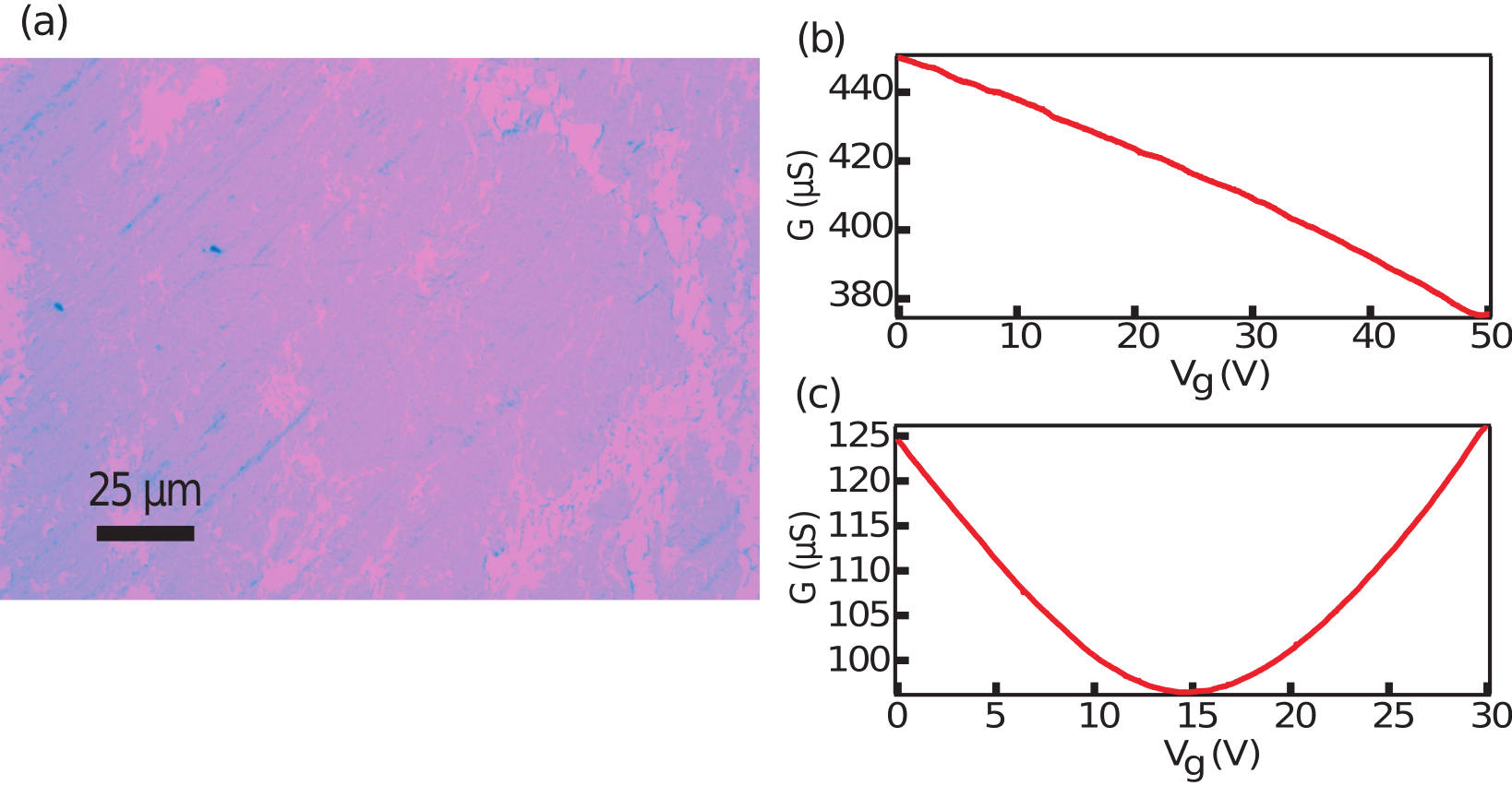}}
\caption{Electrical transport measurements for devices fabricated using graphene samples transferred by two different methods.  (a) Optical image of CVD graphene grown on Cu transferred by SPeeD method. (b) Gating curve for the device fabricated with graphene transferred by conventional Cu etching method. (c) Gating curve for device fabricated with graphene transferred by SPeeD method. The source-drain spacing of the devices used for measurements was $\sim$\SI{5}{\micro\metre}.
}
\label{fig:fig4}
\end{figure}

The Dirac point for graphene transferred by Cu etching is shifted in high positive gate voltage region ($>50$~V) (Figure~\ref{fig:fig4}(b)) indicating that it is highly p-type doped which also agrees well with the Raman measurements. The Dirac point is observable in DI water transferred graphene at $\sim$$15$~V (Figure~\ref{fig:fig4}(c)) indicating that the sample is much cleaner in comparison.

\begin{figure}[!h]
\centerline{\includegraphics[width=90mm]{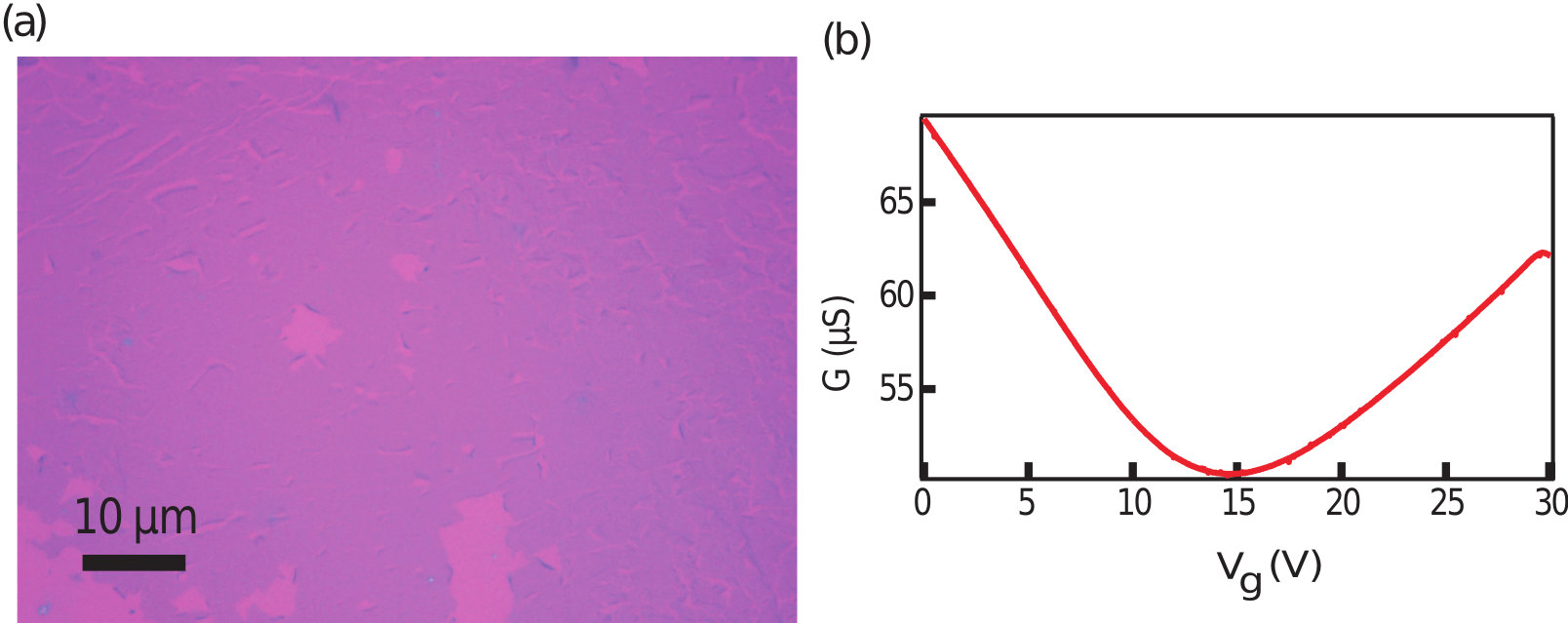}}
\caption{Electrical transport measurement for device fabricated using graphene sample transferred from Pt substrate. (a) Optical image of the graphene transferred by SPeeD from Pt substrate to $300$~nm SiO$_{2}$-coated p-type doped Si substrate. (b) Transport measurement for the device fabricated on the graphene transferred from Pt. The source-drain spacing of the device used for measurement was $\sim$\SI{7}{\micro\metre}.
}
\label{fig:fig5}
\end{figure}

An optical image of the graphene transferred from Pt substrate by SPeeD method on $300$~nm SiO$_{2}$-coated p-type doped silicon target substrate is shown in Figure~\ref{fig:fig5}(a). The gating curve for the device fabricated using our transfer method shows the presence of the Dirac peak at $\sim$$15$~V (Figure~\ref{fig:fig5}(b)) once again showing a relatively clean sample.

In summary, we have successfully demonstrated a novel and simple Soak-and-Peel Delamination method using DI water to transfer CVD-grown graphene from metal substrates like copper and platinum to other substrates of interest. This method does not expose graphene to any harsh chemicals and hence ensures that electrical properties of graphene are not affected. This method is cost effective because no etchant is used and since the metal is not etched it can also be recycled many times reducing large scale production costs. It will also allow the use of single crystals of  Cu(111) for improved growth without consuming the copper single crystals. This simple technique demonstrates low cost, clean transfer of graphene and opens doors for its widespread use. The SPeeD process may also provide a generic route to exploit differential hydrophilic/ hydrophobic interactions to delaminate other 2D layered materials from grown substrates.

\begin{acknowledgements}
The work at TIFR was supported by the Government of India.
\end{acknowledgements}

\end{document}